**Exploring causal relationships in elastomeric fracture**

Asal Y Siavoshani, Zehao Fan, Shi-Qing Wang

School of Polymer Science and Polymer Engineering, University of Akron, Akron, Ohio 44325

## ABSTRACT

The conventional interpretation of elastomeric fracture treats the tearing energy $G$ as an output determined by tear speed, taken to equal crack velocity $v_c$, concluding that higher $v_c$ produced greater viscoelastic dissipation to increase $G$. Based on spatially and temporally resolved polarized optical microscopic (*str*-POM) measurements of tip stress $\sigma_{tip}$, we use prenotched pure shear experiments to clarify the causality: Local crack-tip stress determines the network lifetime, making $v_c$ the passive kinetic output. In the elastic limit that can be readily achieved for a well crosslinked elastomer in a wide range of temperatures, Rivlin-Thomas scaling holds, and $v_c$ depends only on the applied strain, independent of stretch rate, indicating one-to-one correspondence between $v_c$ and $\sigma_{tip}$. In a highly stretchable elastomer made with reduced crosslink density, $v_c$ no longer correlates well with the far-field load as rate-dependent viscoelastic processes emerge to affect network lifetime. Because of the rheological effects on how chain tension builds in the elastomeric network, $v_c$ is higher at a common nominal strain when it is imposed with higher stretch rate. confirm that different stretch rates produce different levels of tip stress. The same $\sigma_{tip}$ still produces the same $v_c$. With *str*-POM measurements we are also able to elucidate the nature of temperature dependence: at a given $\sigma_{tip}$', crack growth is slower at lower temperatures; conversely it requires higher $\sigma_{tip}$ to produce the same $v_c$ at lower temperatures.

## 1.INTRODUCTION

The fracture toughness of covalently cross-linked elastomers varies with both loading rate and ambient temperature[1-4], unlike brittle, strain-intolerant plastics where a constant critical energy release rate can usually capture fracture characteristics.[5, 6] For seven decades this rate and temperature dependence has been described almost entirely through Griffith-Irwin fracture[7-9] mechanics, which assigns[4, 10-12] any rise in fracture resistance to viscoelastic dissipation in the highly stressed zone around the crack tip. Judging the stress and strain fields at a sharply deformed elastomer tip to be intractable and experimentally inaccessible,[7, 13] early work set aside local mechanical criteria in favor of global energy balance. The energy release rate, $G$, was then treated as more than an evaluation parameter and came to be regarded as the physical driver of fracture itself.[9] In this energy perspective, tear speed, taken as equal to the crack propagation velocity $v_c$, was regarded as the independent input, and the larger viscoelastic dissipation it generates raised the tearing energy $G$ as the dependent output[1, 5, 14, 15]. Specifically, such understanding of phenomenology is commonly summarized as

$$G = G_0[1+f(v_c a_T)], \tag{1}$$

where crack speed $v_c$ is considered to change with temperature as the reciprocal of the WLF shift factor $a_T$.

Here we carry out prenotched pure shear experiments that questions this conventional causal picture, represented by eq 1, and demonstrates the reverse dependence: the local crack-tip stress ($\sigma_{\text{tip}}$) sets the network lifetime ($t_{\text{ntw}}$), so that the crack propagation velocity $v_c$ becomes a passive kinetic output governed by thermally activated covalent bond scission.[16-18] Much of the historical confusion over causality traces to the kinematic locking built into classic test geometries such as the trouser tear test of Rivlin and Thomas.[7] In trouser tearing, the crosshead speed is kinematically tied to the crack propagation speed $v_c$ , which led the field to read mechanical resistance as rate-dependent bulk dissipation[14, 15] rather than[16] a local stress-lifetime response.

The prenotched pure shear geometry decouples the crosshead stretching speed from the internal crack velocity $v_c$[19]. For a well-crosslinked elastomer, an elastic limit is reached across a wide temperature range, where Rivlin-Thomas scaling holds strictly and the crack velocity $v_c$ depends only on the applied nominal stretch ratio ($\lambda = h/h_0$), independent of stretching rate. Collapse of the data onto a single master curve fixes a one-to-one correspondence between $v_c$ and the local tip stress $\sigma_{\text{tip}}$, showing that $v_c$ is set by the state of tip stretch rather than by bulk viscous dissipation.

This rate-independent collapse breaks down once the crosslink density is lowered to make the network highly stretchable. There, $v_c$ no longer simply depends on the nominal strain, because rate-dependent viscoelastic (rheological) processes now govern the network lifetime. Since these processes control how chain tension builds up, $v_c$ increases at a fixed nominal strain when the strain is imposed faster. Spatially and temporally resolved polarized optical microscopy (*str*-POM)[20, 21] shows that different stretch rates generate different tip-stress levels at the same nominal strain.

These measurements establish more generally that the same local tip stress $\sigma_{\mathrm{tip}}$ produces the same crack velocity $v_c$ however it is reached, so $\sigma_{\mathrm{tip}}$ remains the controlling variable.

At a fixed tip stress $\sigma_{\mathrm{tip}}$ (equivalently a fixed stress intensity factor $K$), crack growth is systematically slower at lower temperature and faster at higher temperatures. Lower temperature lengthens the lifetime of load-bearing strands under tension, resulting in slower crack propagation. In summary, by comparing modestly stretchable, well-crosslinked networks with highly stretchable, low-crosslinked viscoelastic networks over a wide range of stretch rates, this study reassigns the physical causality of elastomeric fracture: crack speed is governed by local crack-tip stress that determines chain scission kinetics, not by bulk energy dissipation.

## 2. EXPERIMENTAL SECTION

### 2.1 Materials and Network Preparation

The materials investigated in this study included a crosslinked styrene–butadiene rubber (SBR) and crosslinked butadiene rubber (BR).

To eliminate spatial fluctuations in network crosslink density and ensure molecular homogeneity, the crosslinked SBR elastomers were synthesized via a solution-blending, peroxide-initiated crosslinking protocol. A masterbatch was prepared by co-dissolving 100 g of neat SBR and a prescribed quantity of dicumyl peroxide (DCP; 0.1 g for the weakly crosslinked SBR0.1phr sample) in 400 mL of toluene. The solution was subjected to continuous, vigorous stirring for 72 h to ensure a completely homogeneous distribution of the peroxide initiator throughout the precursor matrix.

Following complete dissolution, toluene was systematically removed in a vacuum oven at 70 °C for 24 h until the weight of the dried compound stabilized. The dried, precursor-infused SBR was subsequently compression-molded into isotropic sheets using a heated hydraulic press at 150 °C for 20 min, driving the thermal decomposition of DCP to completion. Specimens were cut from the cured sheets using a custom-machined steel die to ensure defect-free, smooth edges. For uniaxial testing, the dogbone-shaped specimens possessed a gauge length of 17 mm and a nominal width of 1.56 mm, with an average thickness of 1.3 mm and used as a reference. Prenotched specimens for pure shear fracture tests were prepared by introducing an initial edge crack of prescribed length perpendicular to the loading direction using a fresh, sharp razor blade.

Isotropic butadiene rubber (BR) sheets were prepared utilizing DCP as the curing agent. To ensure homogeneity prior to curing, a 100 g batch of neat BR was thoroughly masticated on a two-roll mill for 30 min. To systematically explore the physical effects of network density and chain stretchability, the crosslinked density was controlled by incorporating either 1 wt% (1 phr) DCP to produce highly crosslinked, tight networks, 0.1 wt% (0.1 phr) DCP or 0.03 wt% (0.03 phr) DCP to produce highly stretchable, low-density crosslinked networks. The resulting compounds were preheated at 140 °C for 5 min and compression-molded at 150 °C for 1 h to yield fully cured, isotropic specimens.

2.2 Methods

The mechanical behavior and crack propagation experiments on the SBR and BR samples were carried out using an Instron 5969 universal testing machine equipped with an environmental chamber for precise temperature control. Specimens were secured between two clamps and subjected to both continuous and stepwise tensile loading. During the experiments, the deformation of the samples was visually recorded using a Mokose 4K camera fitted with an Arducam (5–100 mm), zoom lens.

To resolve the spatial and temporal evolution of local stress during deformation, particularly to capture the stress state near the highly concentrated crack tip we employ spatially and temporally resolved polarized optical microscopy (str-POM). This technique exploits the fundamental physical link between molecular orientation and optical anisotropy. In an entangled elastomer, uniaxial or localized deformation stretches the polymer strands, inducing directional orientation of the chain backbones. This molecular alignment generates optical anisotropy, manifesting as a local change in birefringence ($\Delta n$). Within the linear regime of optical anisotropy, this birefringence is directly coupled to the local principal stress difference through the classic stress-optic law:

$$\Delta n = C(\sigma_1 - \sigma_2) \quad (2)$$

where $C$ is the stress-optic coefficient of the elastomer, and $\sigma_1 - \sigma_2$ represents the principal stress difference.

To quantitatively map this optical response in real time, an optical polarization train is integrated directly into our custom tensile stretching apparatus. The setup is configured in a standard transmission arrangement. The specimen is illuminated using a monochromatic sodium vapor lamp or white light as the light source, monochromatic sodium light operates at a clean wavelength of $\lambda = 589$ nm ,to avoid chromatic aberration and ensure well-defined interference fringes. This light is passed through a high-precision linear polarizer before impinging on the specimen. After transmitting through the deformed sample, the light passes through a crossed analyzer.

The resulting optical transmission patterns representing the evolution of stress-induced birefringence and spatial retardation are captured continuously by a high-resolution 4K digital camera synchronized with the mechanical tensile testing unit. By correlating the intensity of the transmitted light and the spatial evolution of the optical fringes with the macroscopic load-deformation data, we directly translated the local optical retardation into a quantitative, real-time map of the local stress field. This high-resolution optical approach allowed us to bypass the ambiguity of remote macroscopic measurements, providing quantitative access to the crack-tip stress that governs fracture kinetics.

## 3. RESULTS AND DISCUSSION

Exploring the causality in elastomeric fracture requires experimental geometry such as pure shear that breaks the kinematic lock built into conventional tear tests. In trouser tearing popularized by Rivlin, Thomas and others, the crosshead stretching speed ($\dot{\lambda}$) is mechanically tied to the crack speed ($v_c \approx \dot{\lambda}$), and the resulting relationship between tear speed and tearing energy $G$[15, 22] has long been read as evidence that a higher crack speed is an independent input driving greater bulk viscoelastic dissipation, and hence a higher $G$.[1, 23] Prenotched pure shear removes this constraint: When physical causality is restored, it becomes transparent that $v_c$ is not a physical driver of bulk dissipation; rather, local crack-tip stress ($\sigma_{\text{tip}}$) determines the network lifetime ($t_{\text{ntw}}$), making $v_c$ the passive kinetic output**.** The far-field load simply acts as a macroscopic control parameter that dictates the level of mechanical stress localized at the notch tip. This local tip stress $\sigma_{\text{tip}}$ governs the lifetime of the network strands via thermally activated covalent bond scission.[16] Thus, local tip stress is the active physical cause, and the observed crack velocity is the passive kinetic effect. Section 3.1 studies a well-crosslinked, modestly stretchable elastomer in its elastic limit, where $v_c$ depends only on the applied strain and collapses onto a single master curve independent of nominal stretch rate, revealing the one-to-one correspondence between $v_c$ and $\sigma_{\text{tip}}$. Section 3.2 turns to a highly stretchable, low-crosslink-density network in which rate-dependent entanglement dynamics make the far-field loading curves rate-dependent in their own right; *str*-POM shows that the same correspondence between $\sigma_{\text{tip}}$ and $v_c$ persists even though the far-field strain-stress curves no longer collapse for different stretching rates. Section 3.3 examines the temperature effect, showing that at a fixed $\sigma_{\text{tip}}$ or stress intensity factor $K$, crack growth is governed by thermally activated bond scission rather than by bulk viscoelastic flow, thus slower at lower temperatures.

### 3.1 Weakly stretchable elastomers

In the elastic limit, reached across a wide range of temperatures and stretch rates for a well-crosslinked elastomer, we study a well-crosslinked styrene-butadiene rubber (SBR, cured with 0.1 phr peroxide) that has been used[18,24] in previous investigations. Because its high crosslink density limits ultimate stretchability, this SBR0.1phr fractures without significant time-dependent notch-tip blunting or macroscopic structural evolution seen in the more stretchable network that is studied in Section 3.2. Rivlin–Thomas scaling holds strictly, and the local crack-tip stress $\sigma_{\text{tip}}$ and hence the level of network stretching in the fracture zone is set uniquely by the instantaneous applied nominal strain ($\lambda = h/h_0$), independent of the stretch rate ($\dot{\lambda}$) at which that strain was reached. Figures 1a-e report crack growth produced at different applied rate. Specifically, Figures 1a and 1b show the stress vs. strain curves and crack speed change during continuous stretching. Figure 1c show that different stretch rates produce the same relationship between $v_c$ and nominal strain, involving two different sample heights $h_0$ = 10 and 20 mm. Figure 1d shows that the same $v_c$ is observed at a given load characterized by $K_{ps} = \sigma\sqrt{h_0}$. Figure 1e shows that $v_c$ is independent of the applied rate and depends solely on the tip stress.

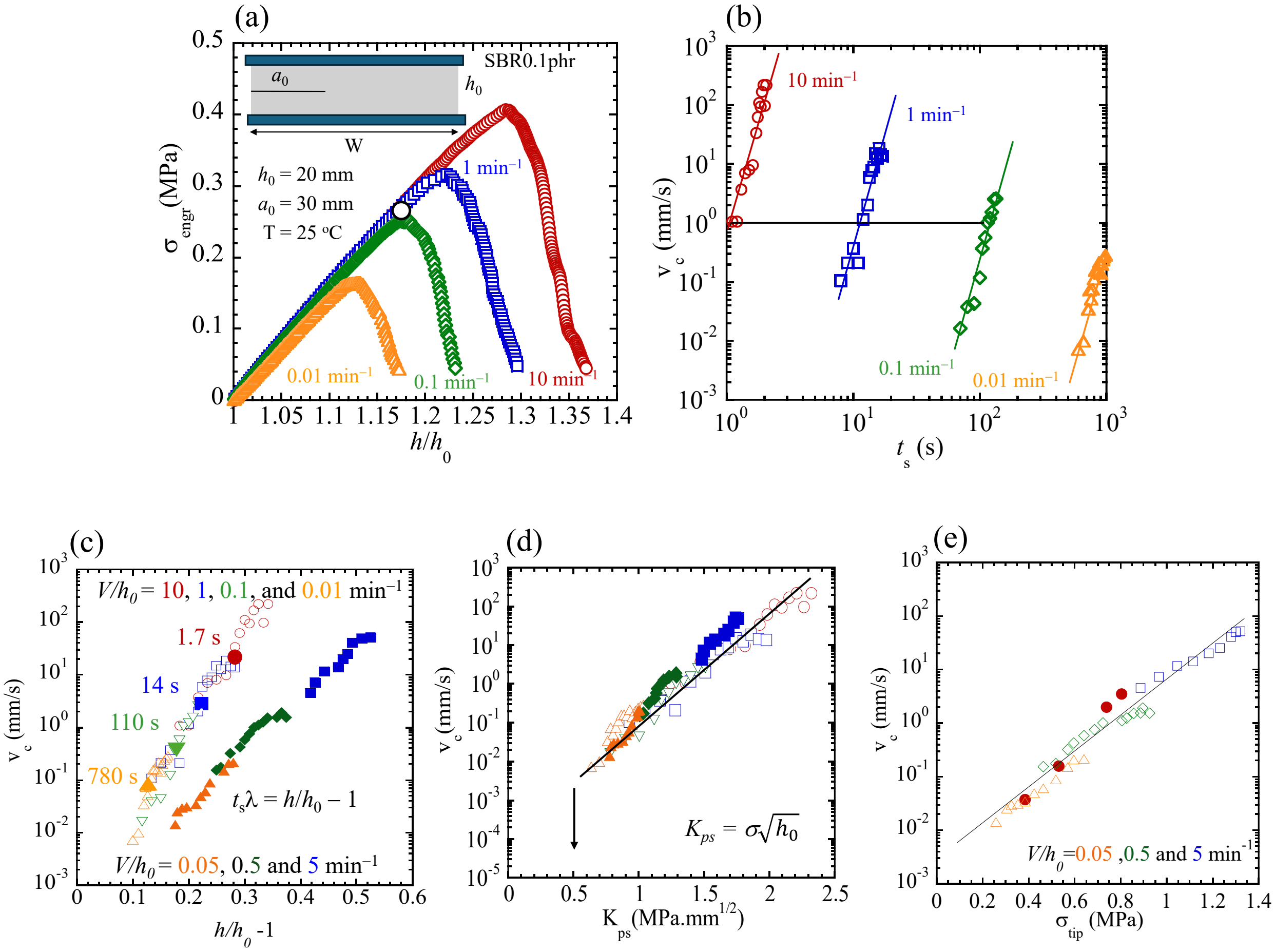

**Figure 1.** Pure-shear fracture behavior of pre-notched styrene–butadiene rubber (SBR0.1phr) at 25 °C. (a) Stress vs. stretch ratio four different rates $\dot{\lambda} = V/h_0 = 0.01, 0.1, 1$ and 10 min$^{-1}$ obtained using specimens of initial heights $h_0 = 20$ mm. (b) Crack growth speed $v_c$ as a function of time, where the horizontal dashed line indicates a common crack speed acquired on different timescales, for $h_0 = 20$ mm. (c) $v_c$ plotted against nominal strain for specimen heights ($h_0 =$ 10 mm and 20 mm). (d) Crack speed $v_c$ plotted against the pure shear stress intensity factor $K_{\rm ps} = \sigma\sqrt{h_0}$ showing a rate-independent collapse for $h_0 = 20$ mm (open symbols) and $h_0 =$ 10 mm (filled symbols). (e) Collapse of all crack speed data onto a single master curve when plotted against the local crack-tip stress $\sigma_{tip}$ resolved via stress-birefringence (str-POM), combining continuous stretching (open symbols, $\dot{\lambda} = 0.05$, 0.5, and 5 min$^{-1}$) and stepwise stretching (filled symbols, halted at different draw ratios during stretching using cross head speed of 0.05 min$^{-1}$) tests, all for $h_0 = 10$ mm.

This residence-time effect is appreciable: at the slowest rate (0.01 min$^{-1}$) a specimen takes roughly a thousand times longer to reach a given nominal strain than at the fastest rate (10 min$^{-1}$), so the crack tip is held under sustained tension long enough for the local molecular "kinetic clock" of bond scission to run to completion even at very low stress producing the exceptionally slow crack speeds captured by the triangles in Figure 1b, where $v_c$ is plotted against elapsed time. At the fastest rate, by contrast, the external loading simply outruns the bond-rupture kinetics at low strain: the specimen is never held there long enough for slow crack growth to register, so the same low-speed regime is invisible in that curve. The horizontal line in Figure 1b marks a single crack speed reached at very different elapsed times across the four rates at the same nominal stress (as indicated by the large open circle in Figure 1a, showing the clearest signature that stress, not stretch rate, sets the lifetime associated with the particular crack speed.

To test whether this stress-driven picture behind the Rivlin-Thomas scaling,[25] we conducted the experiment with two initial heights ($h_0 = 10$ mm and $h_0 = 20$ mm). As shown in Figure 1c, data from each value of $h_0$ form a master. At any given nominal strain, the taller specimen ($h_0 = 20$ mm) drives a systematically higher crack speed $v_c$ than the shorter one ($h_0 = 10$ mm). The $h_0 = 10$ mm data are systematically lower master curve in Figure 1c because there is less stress intensification at notch tip and thus lower $\sigma_{\text{tip}}$ at the same nominal strain. The near-exponential rise of $v_c$ with strain seen in both master curves is itself diagnostic: since network lifetime $t_{\text{ntw}}$ is proportional to $1/v_c$, an exponential dependence of $v_c$ on strain implies that local strand tension reduces the activation barrier for covalent bond scission linearly, as described by the Kinetic Activation Theory of Bond Dissociation (KATBD).[16]

The central result of this section is in Figure 1d: plotted against $K_{\text{ps}}$ instead of the applied nominal strain ($h/h_0$), the $v_c$ data from all four stretch rates for both $h_0$ = 10 mm (filled symbols) and 20 mm (open symbols) specimens collapse onto one super master curve. Because $K_{\text{ps}}$ determines the local stretch state, this rate-independent collapse is direct evidence that $v_c$ and $\sigma_{\text{tip}}$ are in one-to-one correspondence, with no residual dependence on how fast that stretch state was reached. Across five decades of crack velocity the super master curve confirms that $K_{\text{ps}}$, not $h_0$ or $\dot{\lambda}$, sets $v_c$. In other words, instead of eq 1, Figure 1d shows that $v_c$ may be regarded as explicitly dependent on the applied load, e.g., characterized by $K_{\text{ps}}$.

To test whether $\sigma_{tip}$ is indeed the quantity governing crack growth, we carry *str*-POM measurements of tip stress and compare continuous stretching with an entirely separate stepwise protocol (described in detail in Section S1 of the Supplementary Information). In the stepwise tests, the stretching is terminated at a series of fixed stretch ratios λ using $\dot{\lambda} = 0.05$ min$^{-1}$ ($h_o = 10$ mm), and $v_c$ together with $\sigma_{tip}$ resolved by str-POM was recorded under these static conditions. Figure 1e plots $v_c$ against $\sigma_{tip}$ for both data sets together: open symbols correspond to continuous stretching at 0.05, 0.5, and 5 min$^{-1}$, and filled symbols correspond to the stepwise tests at discrete stretch ratios of λ=1.1, 1.2, 1.3, 1.4 and 1.45. Movie S1 shows an example of such a stepwise test involving λ =1.3. All data points overlap to a significant degree. Because this well-crosslinked SBR shows no entanglement effect which means, a given λ delivers the same $\sigma_{tip}$ whether reached continuously or approached in steps and then held; consequently, both protocols produce the same $v_c$ whenever $\sigma_{tip}$ is the same. Figure 1e stresses that $\sigma_{tip}$ alone, not stretch rate, sets the crack speed $v_c$.

In summary, for a well-crosslinked elastomer in its elastic limit, $v_c$ depends only on the applied strain and chosen value of $h_0$ and collapses onto a single rate-independent master curve, revealing direct evidence for a one-to-one correspondence between $v_c$ and $\sigma_{\mathrm{tip}}$. The apparent "toughening" of elastomers at higher loading rates is not a change in fracture resistance at all; it is an artifact of outrunning a finite, stress-dependent network lifetime $t_{\mathrm{ntw}}$, exactly as KATBD predicts. This rate-independent baseline is the benchmark against which the rate-dependent network of Section 3.2 and the temperature-dependent network of Section 3.3 are tested below.

3.2 Highly stretchable elastomers

Having established that a well-crosslinked network operating in the elastic limit yields a rate-independent collapse of $v_c$ against nominal strain (far field load), we now investigate whether there still exists a one-to-one correspondence between tip stress and far-field load for a more stretchable elastomer, e.g., BR with lower crosslink density. Specifically, we study a BR cured with only 0.03 phr of peroxide, stretched in pure shear ($h_0 = 10$ mm) using three stretching rates of $\dot{\lambda} = 0.2, 2,$ and $20\ \mathrm{min}^{-1}$. In this lightly crosslinked network, a dense population of physical entanglements emerges to participate in load bearing, fundamentally altering how chain tension is built up at the notch tip. Figure 2a shows higher nominal stress response in prenotch pure shear for a higher rate. Like Figure 1b, Figure 2b reports the crack speed $v_c$ during stretching at these rates. At a common crack speed, indicated by the dashed line, the corresponding nominal stresses and strains are shown by the three solid symbols in Figure 2a. However, when plotted against the nominal strain, Figure 2c shows $v_c$ to follow separate curves, contrasting sharply with Figure 1c. Moreover, according to *str*-POM measurements of notch tip stress, tip stress grows with engineering stress in a manner that depends on the applied rate. Figure 2d shows that the higher rate produces higher tip stress at comparable nominal stress, where the dashed line marks the same moments indicated by the dashed line in Figure 2b and the solid symbols in Figure 2a. In other words, the different rates produced the same tip stress at different stresses due to the viscoelastic characteristics of this stretchable BR. Although the same nominal stress does not result in the same crack speed, like Figure 1e, independent of stretching rate, $v_c$ still only depends on the tip stress as shown in Figure 2e.

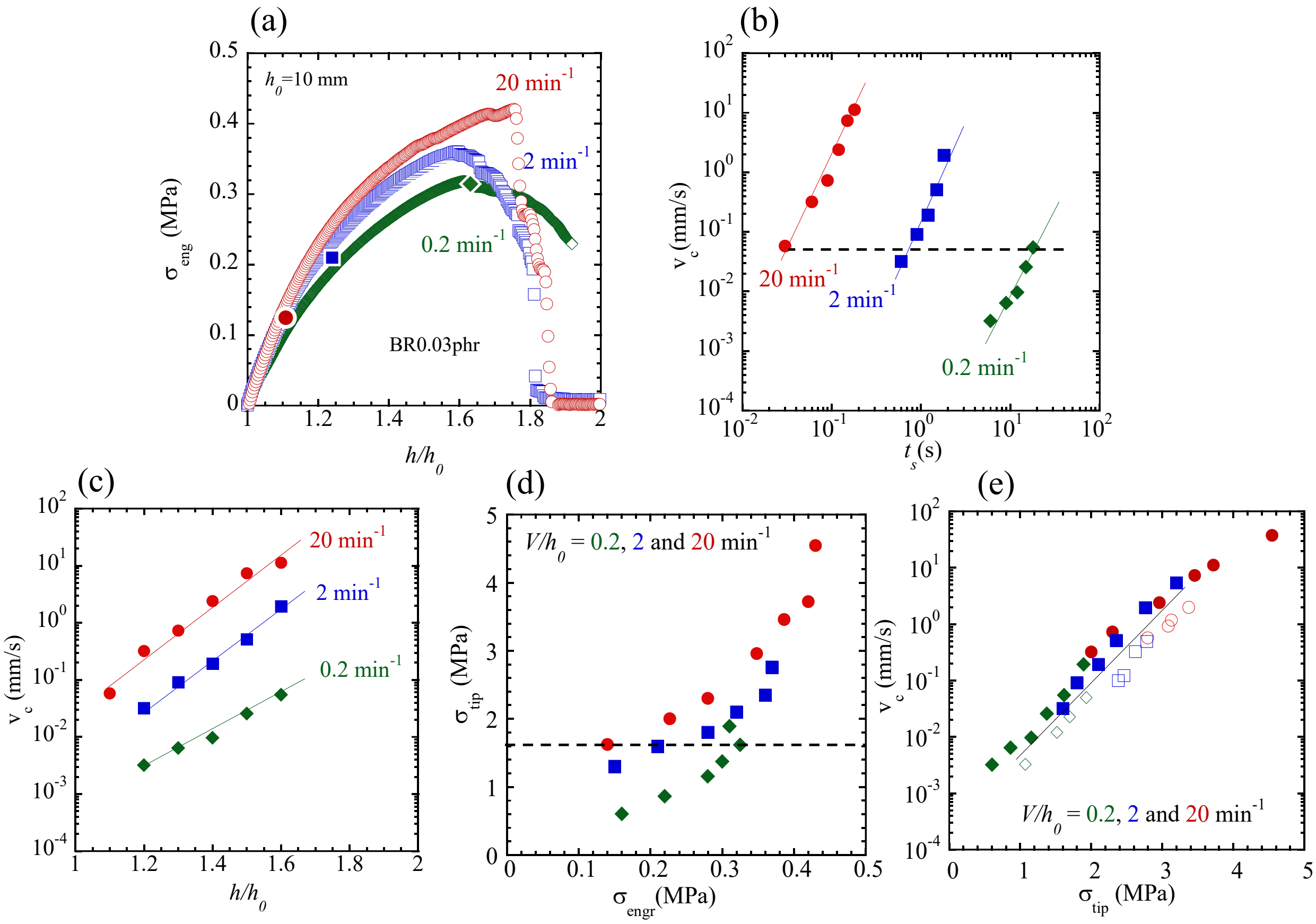

**Figure 2.** Fracture behavior and local notch-tip stress in lightly cross-linked, butadiene rubber (BR0.03phr) tested in pure shear ($h_0 = 10$ mm) at room temperature (25 °C). (a) Nominal stress vs. stretch ratio curves of a prenotched BR0.03phr, continuously stretched at three different rates: 0.2, 2, and 20 $\text{min}^{-1}$. Closed symbols indicate the specific far-field stress states required to sustain identical crack propagation speeds $v_c$ (b) Crack growth speed $v_c$ plotted against time ($t_s$). (c) Crack speed $v_c$ as a function of the nominal stretch ratio $h/h_0$, showing three parallel rate-dependent curves for continuous stretching. (d) local tip stress $\sigma_{tip}$ as a function of nominal far-field stress $\sigma_{\text{engr}}$. (e) Collapse of all crack speed data onto a single master curve when plotted against the local crack-tip stress $\sigma_{tip}$, measured by *str*-POM, from both continuous stretching in filled symbols and stepwise stretching tests in open symbols to four stretch ratios λ =1.2 ,1.3 1.4 and 1.45.

In BR0.03phr where there is plausibly embedded chain uncrossability (leading to physical entanglement), considerably higher network stretching at notch tip can occur at higher rates although the far-field stress shows weaker rate dependence. For example, when stretched rapidly, more load-bearing strands may emerge due to entanglement so that the overall higher chain stretching leads to shortened network lifetime, leading to fast crack growth.

To verify the universal relationship between $v_c$ and $\sigma_{tip}$, we also carry out stepwise stretching to four stretch ratios of $\lambda$ =1.2 ,1.3 1.4 and 1.45 using three different rates of 0.2, 2 and 20 min$^{-1}$. See details in Section S2 of the Supplementary Information. When higher rate is used there is higher transient tip stress $\sigma_{tip}$, producing higher $v_c$, as shown by the open symbols in Figure 2e. Over time after stepwise stretching, tip stress relaxes, and $v_c$ decreases accordingly, as shown, for example, by Movie S2 in SI, involving stepwise stretch to λ =1.3. The open symbols in Figure 2e are measurements of $v_c$ right after stepwise stretch, where three symbols represent the three rates of 0.2 (diamonds), 2 (squares) and 20 (circles) min$^{-1}$, where the first, right-most symbols correspond to λ =1.45 and the left-most symbols correspond to λ = 1.2.

3.3 Temperature effect

After demonstrating what controls crack growth in elastomers in the preceding two subsections, we turn to clarification of the origin of the temperature effect on crack growth. For seven decades, Greensmith-Thomas landmark paper[1] have formed the basis for elastomeric fracture. However, the tear tests have continued to blur the underlying causality and confuse the investigation regarding the origin of the temperature dependence. With well crosslinked BR1phr, it suffices to adopt single-edge notch (SEN) and observe how temperature affects the magnitude of $v_c$. From a recent study[24] on BR1phr, we can extract information on the crack speed $v_c$ during continuous stretching in SEN at different temperatures.

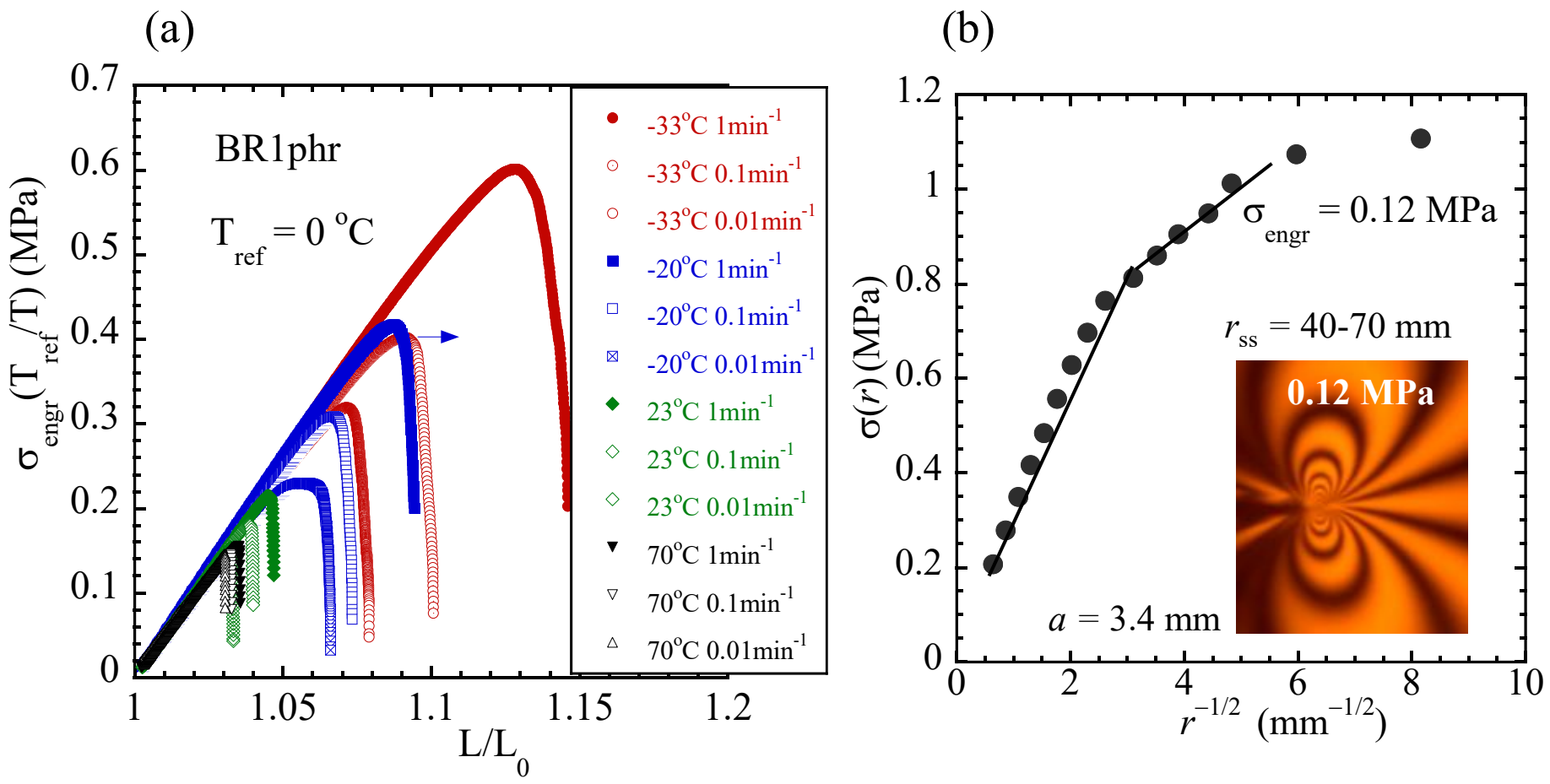

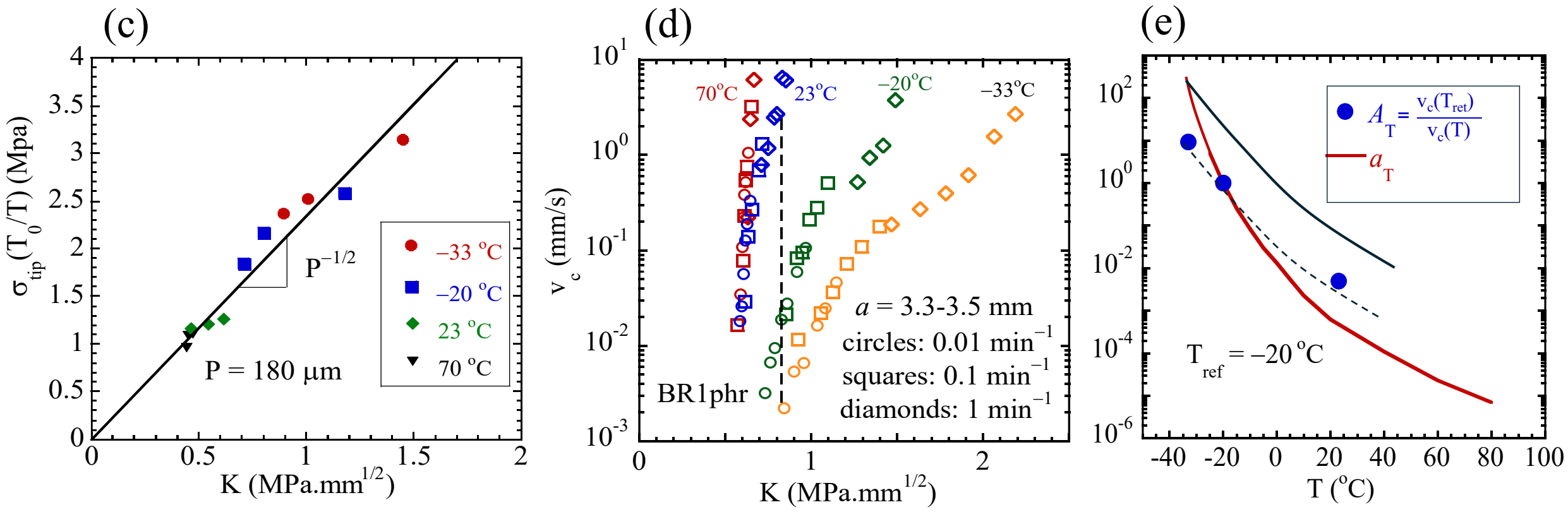

**Figure 3.** Temperature dependence of fracture and crack-tip stress in highly cross-linked butadiene rubber (BR1phr) tested in a single-edge notch (SEN) geometry. (a) Nominal stress vs. stretch ratio at different temperatures using $\dot{\lambda}$ = 0.1 min$^{-1}$ and $L_0$ = 30 mm. (b) *Str*-POM measurements at crack tip, revealing a finite stress saturation zone (SSZ) of $r_{ss} \approx$ 40–70 μm during crack propagation at nominal far-field stress $\sigma_{engr}$ = 0.12 MPa. (c) Linear relationship between the experimentally resolved crack-tip stress $\sigma_{tip}$ and stress intensity factor $K$. (d) Crack speed $v_c$ plotted against $K$ at different temperatures; different symbols representing various stretch rates overlap at each temperature, and the vertical dashed line at $K$ = 0.82 MPa. (e) Normalized reciprocal crack velocity $v_c$ as a function of temperature shown by the three circles where the upper thinner curve is the line through the three circles after any arbitrary vertical shift of the dashed curve for comparison with the WLF curve ($a_T$).

Figure 3a shows the nominal stress versus stretch ratio curves of BR1phr in SEN at different temperatures. While Sections 3.1 and 3.2 utilized pure shear, results based on SEN tests in Figures 3a-d reach the same fundamental conclusion. Here, BR is highly crosslinked and therefore largely a linear elastic elastomer, making SEN a valid configuration for LEFM analysis where the far-field load dictates the tip stress buildup[16] through the stress intensity factor $K = \sigma_{engr}\sqrt{\pi a}$.

To resolve the local stress at notch tip produced by the far field stress, *str*-POM measurements were carried out. Figure 3b shows how the stress field at the notch tip increases during stretching with crosshead speed of $\dot{\lambda}$ = 0.1 min$^{-1}$ Specifically, a finite size $r_{ss}$ (≈ 40–70 $\mu$m) of the stress saturation zone (SSZ) is observed during crack growth.

Figure 3c shows a linear relationship between local tip stress and $K$: higher values of the driving force $K$ produce higher tip stresses to result in higher $v_c$. Specifically, in agreement with the pure-shear findings of Section 3.1, at each temperature $v_c$ depends only on $K$, independent of the stretch rates, represented by different symbols, used to produce the far-field stretching, as demonstrated in Figure 3d by the overlapping at each of three temperatures. Reading vertically along the dashed line at a common value of $K = 0.82$ MPa $\cdot$ mm$^{1/2}$ across the three temperatures, we see that $v_c$ increases systematically with temperature, reflecting the temperature dependence of network lifetime at the crack tip. It is also worth noting from Figure 3d, reading the data horizontally, that at lower temperatures higher loads are required to maintain the same speed $v_c$.

Finally, Figure 3e sheds light on the molecular origin of this temperature dependence. Specifically, the dependence of crack speed $v_c$ on temperature is weaker than that of the WLF shift factor $a_{\mathrm{T}}$, which is by the lower curve. This discrepancy between $a_{\mathrm{T}}$ and reciprocal of $v_c$ points to a different origin of temperature dependence in elastomeric fracture, associated with the thermally activated lifetime of stretched network, consistent with the assertion that $v_c$ is the passive kinetic output of local crack-tip stress.

## 4.CONCLUSIONS

Griffith-Rivlin-Thomas energy-balance perspective[1, 7] treated the energy-release-rate or tearing energy $G$ as output, while regarding tear speed (equated with crack velocity $v_c$) as an independent input that produces viscoelastic dissipation, to be read as the origin of $G$. To clarify the causal relationship between $G$ and $v_{c,}$ prenotch pure shear experiments are carried out to decouple the crosshead stretching kinematics from the internal crack motion and suggest to reverse the previously proposed causal chain in eq 1, as shown in Figure 1e and Figure 2e: local tip stress $\sigma_{\mathrm{tip}}$ sets the lifetime of the load-bearing strands at the crack plane through thermally activated bond scission, and the crack velocity $v_c$, related by the reciprocal of that lifetime, is the passive kinetic output of $\sigma_{\mathrm{tip}}$. Much of the historical confusion traces to a specific artifact of testing geometry: in the trouser tear test that established the field's founding data set, the crosshead speed is kinematically locked to $v_c$ so crack speed and macroscopic loading rate were never separable quantities. Pure shear removes that lock. Once $\sigma_{\mathrm{tip}}$ and $v_c$ can be measured independently, $G$ is exposed as a phenomenological bookkeeping quantity whose value follows from the underlying stress-lifetime process at the tip, rather than a viscous energy dissipation mechanism that drives it.

For a well-crosslinked elastomer held within its elastic limit, this causal structure collapses onto a single quantitative statement. Rivlin-Thomas scaling holds strictly in this regime, $v_c$ depends only on the applied nominal strain, and it is independent of the stretch rate at which that strain is reached, so nominal strain and $\sigma_{tip}$ carry the same information about the local state of the network and their one-to-one correspondence with $v_c$ is direct rather than inferred. The near-exponential rise of $v_c$ with strain documented in Section 3.1 is itself a mechanistic signature: it indicates that strand tension lowers the activation barrier for covalent bond dissociation approximately linearly, as the KATBD[16] describes. According to the stress perspective, the apparent toughening of elastomers observed at higher loading rates in conventional tests is the experimental signature of finite, stress-dependent network lifetime $t_{\mathrm{ntw}}$: faster crack growth must involve shorter lifetime, which is only possible if the elastomer is more stretching, corresponding to higher $G$.

The causal picture extends beyond this elastic limit. In a highly stretchable, low-crosslink-density network, entanglements enter as transient load-bearing elements, and the relation between far-field load and $\sigma_{\text{tip}}$ becomes rate-dependent: faster stretching engages more entanglement-mediated strands and raises $\sigma_{\text{tip}}$ at a fixed nominal stress, so $v_c$ no longer collapses against strain or engineering stress. What stays invariant is the relation between $\sigma_{\text{tip}}$ and $v_c$ itself. *Str*-POM measurements show that different stretch rates produce different levels of tip stress at the same nominal strain, and that the same $\sigma_{\text{tip}}$ produces the same $v_c$ however that stress state was reached, continuously or stepwise. The present data reveal local stress as the physical driver, identifying rate dependence of the buildup of chain tension in the network and shows that once the tension is expressed as a local stress, the kinetics of bond dissociation at the tip proceed on their own terms, uncoupled from the loading history that produced them.

The temperature dependence of fracture submits to the same physics. At a fixed $\sigma_{\text{tip}}$, or equivalently a fixed stress intensity factor $K$, crack growth is systematically slower at lower temperature, supporting the notion that crack growth is thermally activated. This temperature dependence is measurably weaker than that of the Williams–Landel–Ferry (WLF) shift factor that governs polymer relaxation. Thus the temperature dependence of $v_c$ cannot be attributed to the viscoelastic friction that sets the bulk mechanical response. A distinct, more local activation process, the thermally assisted scission of a stretched covalent bond, is involved in crack growth.

Reassigning causality in this way changes what the tearing energy $G$, and the rate and temperature dependence historically read through it, mean. $G$ remains a valid measure of the aftermath but its magnitude follows from the local stress-lifetime process at the tip integrated over the loading history. This distinction carries a practical consequence for elastomer design. Toughening a network by promoting more bulk viscoelastic loss, the strategy implicit in the past seven decades of energy-based fracture mechanics, is a different design lever than toughening a network by slowing the rate at which $\sigma_{\text{tip}}$ quicken network lifetime. The present study establishes this local causal framework across two crosslink densities, three elastomers, and a wide temperature range accessible by pure shear and single-edge-notch testing. Extending it to filled and reinforced elastomers, to fatigue and cyclic loading, and to crack growth under multiaxial or dynamic fields remains open, and $\sigma_{\text{tip}}$ resolved by *str*-POM, or other equivalent local probe, should be the natural variable through which that extension is pursued.

Taken together, these results restore local mechanics to the center of elastomeric fracture. Crack velocity is the outcome of a stress-activated molecular process at the tip, and the energy balance long used to characterize elastomeric fracture is best understood as a phenomenological bookkeeping of this local stress-lifetime relationship. Any future account of rate and temperature dependence in elastomer fracture should be built on that local, thermally activated basis rather than on the global energy balance that has previously organized the field since Griffith, Rivlin, and Thomas.

**ASSOCIATED CONTENT**

**Supporting Information**

The Supporting Information is available free of charge online. Experimental setup details, Movie S1, Movie S2.

## ACKNOWLEDGMENTS

This work is supported, in part, by the Polymers program through Special Creativity Extension of the US National Science Foundation grant DMR-2210184.

**AUTHOR INFORMATION**

**Corresponding Author**

*Shi-Qing Wang – School of Polymer Science and Polymer Engineering, University of Akron, Akron, Ohio 44325, United States;

Email: swang@uakron.edu

**Authors**

Asal Y. Siavoshani – School of Polymer Science and Polymer Engineering, University of Akron, Akron, Ohio 44325, United States

Zehao Fan – School of Polymer Science and Polymer Engineering, University of Akron, Akron, Ohio 44325, United States

**Author Contributions**

The manuscript was written through contributions of all authors. A.Y.S. conducted the investigation, visualization, methodology, data curation, formal analysis, wrote the original draft, and prepared the Supporting Information. Z.F. contributed to investigation, visualization, methodology, data curation, and formal analysis. S.-Q.W. conceptualized the project, contributed to investigation, formal analysis, validation, supervised the research, and revised and edited the manuscript. All authors have given approval to the final version of the manuscript.

**Notes**

The authors declare no competing financial interest.

## REFERENCES

1. Greensmith, H. W.; Thomas, A. Rupture of rubber. III. Determination of tear properties. *J. Polym. Sci.* **1955,** 18, (88), 189-200.
2. Greensmith, H. Rupture of rubber. VII. Effect of rate of extension in tensile tests. *J. Appl. Polym. Sci.* **1960,** 3, (8), 175-182.
3. Smith, T. L. Ultimate tensile properties of elastomers. I. Characterization by a time and temperature independent failure envelope. *Journal of Polymer Science Part A: General Papers* **1963,** 1, (12), 3597-3615.
4. Gent, A. Adhesion and strength of viscoelastic solids. Is there a relationship between adhesion and bulk properties? *Langmuir* **1996,** 12, (19), 4492-4496.
5. Kinloch, A. J.; Young, R. J., *Fracture behaviour of polymers*. Springer Science & Business Media: **2013**.
6. Williams, J. G., Fracture mechanics. In *The Physics of Glassy Polymers*, Haward, R. N.; Young, R. J., Eds. Springer Netherlands: 1997; pp 343-362.
7. Rivlin, R.; Thomas, A. G. Rupture of rubber. I. Characteristic energy for tearing. *J. Polym.Sci.* **1953,** 10, (3), 291-318.
8. Irwin, G. R. Analysis of stresses and strains near the end of a crack transversing a plate. *Trans. ASME, Ser. E, J. Appl. Mech.* **1957,** 24, 361-364.
9. Griffith, A. A. VI. The phenomena of rupture and flow in solids. *Philosophical transactions of the royal society of london. Series A, containing papers of a mathematical or physical character* **1921,** 221, (582-593), 163-198.
10. Persson, B.; Albohr, O.; Heinrich, G.; Ueba, H. Crack propagation in rubber-like materials. *J. Phys.: Condens. Matter* **2005,** 17, (44), R1071.
11. Knauss, W. G. A review of fracture in viscoelastic materials. *Int J Fract* **2015,** 196, (1-2), 99-146.
12. Creton, C.; Ciccotti, M. Fracture and adhesion of soft materials: a review. *Rep. Prog. Phys.* **2016,** 79, (4), 046601.
13. Griffith, A. A. The theory of rupture. *Proceedings of the First International Congress for Applied Mechanics, Delft* **1924**, (Biezeno, C. B; Burgers, J. M. .), 55.
14. Thomas, A. The development of fracture mechanics for elastomers. *Rubber Chem. Technol.* **1994,** 67, (3), 50-67.
15. Tsunoda, K.; Busfield, J.; Davies, C.; Thomas, A. Effect of materials variables on the tear behaviour of a non-crystallising elastomer. *Journal of Materials Science* **2000,** 35, (20), 5187-5198.
16. Wang, S.-Q.; Fan, Z.; Gupta, C.; Siavoshani, A.; Smith, T. Fracture behavior of polymers in plastic and elastomeric states. *Macromolecules* **2024,** 57, (9), 3875-3900.
17. Asal Siavoshani, Z., Muxuan Yang, Shan Liu, Ming-Chi Wang, Jiabin Liu; Weinan Xu, Junpeng Wang, Shaoting Lin and Shi-Qing Wang. How rate, temperature and solvent exchange affect polymer network rupture? *Soft Matter* **2024,** 20, 7657-7667.
18. Siavoshani, A. Y.; Wang, M.-C.; Liang, C.; Jaisingh, A.; Wang, J.; Wang, C.; Wang, S.-Q. Exploring the theoretical foundation with rupture and delayed rupture experiments. *Macromolecules* **2026,** 59, 2885.
19. Thomas, A. Rupture of rubber. II. The strain concentration at an incision. *J. Polym. Sci.* **1955,** 18, 177-188.

20. Smith, T.; Gupta, C.; Fan, Z.; Brust, G. J.; Vogelsong, R.; Carr, C.; Wang, S.-Q. Toughness arising from inherent strength of polymers. *Extreme Mechanics Letters* **2022,** 56, 101819.
21. Wang, S.-Q.; Fan, Z.; Siavoshani, A.; Wang, M.-c.; Wang, J. Fresh considerations regarding time-dependent elastomeric fracture. *Extreme Mechanics Letters* **2025,** 74, 102277.
22. Kadir, A.; Thomas, A. Tear behavior of rubbers over a wide range of rates. *Rubber Chem. Technol.* **1981,** 54, (1), 15-23.
23. Gent, A.; Lai, S.; Nah, C.; Wang, C. Viscoelastic effects in cutting and tearing rubber. *Rubber Chem. Technol.* **1994,** 67, (4), 610-618.
24. Wang, S.-Q.; Fan, Z. Investigating the Dependence of Elastomeric Fracture on Temperature and Rate. *Rubber Chem. Technol.* **2023,** 96, (4), 530-550.
25. Fan, Z.; Wang, S.-Q. Resolving stress state at crack tip to elucidate nature of elastomeric fracture. *Extreme Mechanics Letters* **2023,** 61, 101986.